\documentclass[aps,preprint,groupedaddress]{revtex4-1}

\usepackage[american]{babel}
\usepackage{bm}
\usepackage{amsfonts}
\usepackage{graphicx}
\usepackage{amsmath}
\newcommand{\nn}{\nonumber}
\newcommand{\bs}[1]{\boldsymbol{#1}}

\newcommand{\zd}{\delta}

\newcommand{\zs}{\sigma}

\newcommand{\ze}{\varepsilon}

\newcommand{\zg}{\gamma}

\newcommand{\zm}{\mu}

\newcommand{\za}{\alpha}
\newcommand{\zy}{\psi}

\newcommand{\dpp}[2]{\frac{  \partial #1 }{  \partial #2   } }
\begin{document}
\begin{center}
\Large{\textbf{Quantum atomistic approach for interacting spins}}\\
\vskip0.5cm
	\small{O. Morandi}	\\
\vskip0.5cm
	\textit{\textsf{Institut de Physique et Chimie des Mat\'eriaux de Strasbourg \\
23, rue du Loess,
F-67034 Strasbourg, France}}	
\vskip0.5cm
	\textit{omar.morandi@ipcms.unistra.fr}
\end{center}
\begin{center}
\begin{minipage}[h]{0.8\textwidth}
\section*{\textsf{Abstract}}
 \small
\textsf{The phenomenological Landau theory of the spin precession has been used to reproduce the out-of-equilibrium properties of many magnetic systems. However, such an approach suffers from some serious limitations. The main reason is that the spin and the angular momentum of the atoms are described by the classical theory of the angular momentum. We derive a discrete model that extends the Landau theory to the quantum mechanical framework. Our approach is based on the application  of the quantification procedure to the classical hamiltonian of an array of interacting spins. The connection with the classical dynamics is discussed. }
\end{minipage}
\end{center}
\vspace{0.5cm}
\normalsize

\section{Introduction}

The study of the  magnetism dynamics in a nanostructure is a rapidly developing area of research. The technological interest relies on the applications to magnetic data storage and sensing devices. The conventional way to record information in a magnetic memory is to modify the local magnetization of a magnetic material \cite{Bigot_13}.
Understanding the spin dynamics in magnetic materials is an issue of crucial importance for progress in information processing and recording technology.
As an example of new electronic devices that are very promising for the future applications to the  magneto-electronic technology we cite the Diluted Magnetic Semiconductors \cite{morandi_NJP_09,morandi_PRB_10b,Shen_14,morandi_PRB_10,morandi_PRB_11} and the Rashba diodes \cite{Barletti_10}. For their unique interplay of spin and electronic degrees of freedom these systems are of great interest. They open the possibility to tailor new spintronic devices \cite{Barletti_15}.

The use of a laser field has been proved to lead to a considerable speed up of the digital information recording. The elementary mechanisms that leads to the modification of the magnetic order triggered by laser pulses are at the center of an intense debate \cite{Hinschberger_15,morandi_PRE_14,Baral_14,Manchon_12}.

One of the most successful attempts to reproduce the magnetization dynamics is the so called atomistic approach. The main assumption of the atomistic models is that the atoms of magnetic materials are well described by a local magnetic moment. The first model where a magnetic solid was approximated by an array of interacting localized magnetic moments was proposed by Ising in 1925 for explaining the phase transition of a ferromagnet. In the Ising approach the spin of all the atoms was assumed to be always directed along a fixed direction (collinear spin model). A natural extension of the Ising model was proposed by Heisenberg that considered the atomic spin as a three-dimensional vector. The Heisenberg model allows to describe non collinear effects like for example the spin precession, the magnons, and the spin torque effect. More recently, atomistic models have received renewed attentions \cite{Evans_14}. They are based on the classical description of the atomic spin and angular momentum provided by the phenomenological Landau-Lifshitz-Gilbert equation.
They have been used to reproduce the ultrafast evolution of the total  spin in systems excited by laser pulses and the angular momentum  transfer between metals and rare earth compounds \cite{Bergeard_14}.

However, such models suffer of a serious theoretical limitation that reduces considerably their area of application. The reason is that such atomistic models have been always derived on the basis of the classical spin theory. It is well known that electron spin is intrinsically a quantum mechanical phenomena and the classical description is valid only in the limit of large spin.

In the present contribution, we propose an atomistic-like approach based on the quantum mechanical framework. A cluster of atoms or, more generally, a nanomaterial is described in terms of a network of quantum spin particles interacting via exchange interaction.
Our model is able to describe fully coherent phenomena such as the exchange of spin of angular momentum among the atoms and the effect of the deformation of the atomic orbital induced by the lattice field. The coupling with the phonon bath is modeled by a stochastic approach.

\section{Atomistic spin model: classical description}

The physical properties of a magnetic system arise from the microscopic configuration of the angular momentum and the spin of the atoms.
The so called atomistic approach is a simple and powerful approach that allows the simulation of the microscopic spin dynamics of a complex system. In this framework, a solid is modeled by an array of localized interacting atoms. %In the following, we will discuss the application of such an approach to the study of the dynamics of the magnetism in solid by assuming that the atoms interaction is described by the rules of the classical or the quantum mechanical physics.

In the framework of the classical theory of the magnetism, the magnetic moment $\mu$ of an atom can take any value. However, according to the quantum mechanical description of the spin, the total spin of any system is equal only to some well defined values. It is either an integer or a semi-integer multiple of the Plank constant $\hbar$. The magnetic moment of a quantum particle takes the form
$$
\zm =  - \zg s \hbar\;, 
$$
where either $s=1,2,3,\ldots$ (boson particle)  or $s=\frac{1}{2},\frac{3}{2},\frac{5}{2},\ldots$ (fermion particle). We denote by $\zg$ the giromagnetic ratio $  \frac{g \mu_B}{\hbar }=\frac{|q| g}{2m_e }$, where $\mu_B$ is the Bohr magnetron, $g$ the Lande factor, $m_e$ the electron mass and $q$ the electron charge.

In this section, we derive the evolution equation of a network of interacting magnetic atoms in the framework of the classical mechanics. The total angular momentum of the atoms is modeled by a three-dimensional vector  $\mathbf{S} = s \hbar \widehat{\mathbf{S}}$ (the hat denotes the unitary vector). The magnetic moment is $ {\bs\zm} =  - \zg \mathbf{S}
$. As a first step, we consider a very simple situation. We discard the interactions of the atoms with the environment (i. e. the atom-phonon collision, interaction with a radiative field). In this case the evolution of the classical moment is the simple precession around the local magnetic field $\mathbf{B }$ seen by the atom
\begin{align}
   \dpp{\mathbf{S}^i}{t} =-\zg  \; \mathbf{S}^i \times  \mathbf{B }(\mathbf{R}^i)\;. \label{evol S cons}
\end{align}
Here, we used the notation $\mathbf{S}^i \equiv  \mathbf{S}(\mathbf{R}^i)$ where $\mathbf{R}^i$ is the position of the $i-$th atom.
Inside a solid, the microscopic magnetic field is mainly originated by two phenomenon: the spin-spin interaction of two neighbor atoms (exchange field) and the molecular anisotropy (anisotropy field).
\begin{align}
  \mathbf{B}(\mathbf{R}^i) = \underbrace{ - \frac{1}{  s_i  g\mu_B \hbar  } \sum_{j\neq i, j\in \textrm{ NN}_i } \frac{1}{  s_j  } J_{ij}   \mathbf{S}^j}_{\textrm{Exchange field}}\; \underbrace{-   \frac{2 D}{  \hbar } \frac{ S^i_z }{s_i^2 } \widehat{\mathbf{e}}_z}_{\textrm{Anisotropy field}}+    \mathbf{B}_{0}(\mathbf{R}^i) \label{B}
\end{align}
The symbol NN$_i$ indicates that the sum in the exchange term is taken over the neighbors atoms of the $i-$th atom (usually only the first and the second neighbors are considered). The exchange field  interaction is usually very strong in magnetic systems. It was firstly introduced by Heisenberg and it is very popular for modeling the magnetic interaction between localized atoms. Mathematical models based on the Heisenberg nearest neighbor interaction have been successfully applied to various systems. They are able to reproduce correctly the static and dynamical properties of the magnetism in solids and molecular systems. The anisotropy term is an effective magnetic field that takes into account the interaction of the localized atomic orbitals with the electrostatic lattice field. Such an interaction is described by a nonlinear magnetic field that is quadratic with respect to the projection of the atom spin along a constant direction. For simplicity, we assume that the molecular anisotropy field is directed along to the $z$ axis (we denote the $z-$direction by the unitary vector $\widehat{\mathbf{e}}_z$). The  deformation constant $D$  is a parameter that quantifies  the strength of the anisotropy field. Finally, $ \mathbf{B}_{0}(\mathbf{R}^i)$ denotes some external magnetic field. According to Eq. \eqref{evol S cons} and Eq. \eqref{B} the magnetic system is described by an array  of classical interacting particles for which the total Hamiltonian  is
\begin{align}
 \mathcal{H}_{cl} =- \sum_i \left[ \mathbf{S}^i \cdot\left(   - \frac{1}{  s_i( g\mu_B)^2  } \sum_{j\neq i, j\in \textrm{ NN}_i } \frac{1}{  s_j  } J_{ij}   \mathbf{S}^j +    \mathbf{B}_{0}(\mathbf{R}^i) \right)+ \frac{  D}{ s_i^2 g\mu_B  }     {S^i_z}^2\right] \;. \label{Ham_cl}
\end{align}
It is easy to verify that the magnetic field can be obtained  by the usual expression  
\begin{align}
  \mathbf{B}(\mathbf{R}^i) =  \zg \dpp{ \mathcal{H}_{cl}}{\mathbf{S}^i}\;. 
\end{align}

\subsubsection{Dissipative effects: LLG equation}

It is easy to see that Eq. \eqref{evol S cons} conserves the total energy of the atoms $E=\zg \sum_i \mathbf{S}^i\cdot \mathbf{B}(\mathbf{R}^i)$. In order to apply our model to a real situation, it is necessary to include some dissipative effects. One of the most simple way to include the dissipation of spin and angular momentum in Eq. \eqref{evol S cons} is provided by the Landau-Lifshitz-Gilbert theory. Landau and, independently, Gilbert suggested to modify the equation that describes the precession of the angular momentum around the external field $\mathbf{B}$ by adding a phenomenological term that describes the local alignment of the atomic angular momentum to $\mathbf{B}$. The Landau-Lifshitz-Gilbert (LLG) equation takes the form
\begin{align}
  \dpp{\mathbf{ {S}}^i}{t}= - \frac{\zg}{ (1+\za^2_i) }  \mathbf{S} ^i \times \left(  \mathbf{B}^i - \frac{\za_i}{s_i} \mathbf{S} ^i \times  \mathbf{B}^i \right)\;. \label{landau eq CL}
\end{align}
Here, $\za_i$ is the so called Landau dumping parameter. It measures the strength of the interaction of the atoms with some external thermal bath. The Landau equation models the loss of magnetic energy in a very general manner. What is the physical origin and the source of the dissipation of energy and angular momentum is not specified. The model describes only the transfer of energy and angular momentum to some external bath. In Eq. \eqref{landau eq CL} the value of the temperature of such an external bath is not given. For this reason, it is clear that the LLG equation cannot be able to reproduce the correct atomic spin configuration at the thermal equilibrium.
%In order to model correctly the interaction of the spin system with the external bath, we describe the spin-bath interaction  in terms of a stochastic signal that models the thermal fluctuations of the total magnetic field.

A simple way to model the relaxation of the spin system towards the thermal equilibrium is to consider a stochastic approach. According to Langevin theory, the transfer of energy between the atomic system and the external bath can be described by some stochastic magnetic signal. It reproduces the thermal fluctuations of the total magnetic field. Such a noise signal is modeled by a white distribution ${\bs \xi}^i \equiv {\bs \xi} (\mathbf{R}^i)$. The statistical properties of ${\bs \xi}^i $ are  described by the following expression \cite{Skubic_14}
\begin{align}
\langle\langle {\bs \xi}^i (t){\bs \xi}^j (t') \rangle\rangle & = \zd(t-t') \zd_{ij} 2 \frac{\za_i}{\zg}  k_B T_i \mu_i\;, 
\end{align}
where the bracket denotes the stochastic mean over the statistical ensemble, $T_i$ is the bath temperature, $\zd(t-t')$ and $\zd_{ij}$ are respectively the Dirac and the Kroneker delta. By adding the stochastic distribution ${\bs \xi}$  to the total magnetic field we obtain the complete LLG-Langevin equation \cite{Evans_14}
\begin{align}
  \dpp{\mathbf{ {S}}^i}{t}= - \frac{\zg}{ (1+\za^2_i) }  \mathbf{S} ^i \times \left(  \mathbf{B}^i +{\bs \xi}^i  - \frac{\za_i}{s_i} \mathbf{S} ^i \times  \mathbf{B}^i \right)\;. \label{landau eq CLnoise}
\end{align}
In order to compact the notation we have defined  $ \mathbf{B}^i \equiv \mathbf{B}  (\mathbf{R}^i)$.
\section{Quantum atomistic spin model}
We pass now to the description of the spin system in the quantum mechanical framework. We proceed as follows: at first, we derive the quantum model in the presence of the external field $\mathbf{B}_0 $. As a second step, we include the internal fields (exchange and anisotropy fields). Finally, we consider the dissipative effects.

In the quantum framework, the spin of one atom is represented by a wave function $\psi $ defined in the spinorial  Hilbert space. The quantum mechanical Hamiltonian that describes the magnetic energy of a quantum  particle with total spin $s$ in the presence of the external magnetic field $ \mathbf{B}_0$ is given by the Zeeman expression
\begin{align}
 \mathbb{H}_0 &= \frac{g \zm_B}{\hbar} \mathbf{B}_0 \cdot  \mathbb{S}^i
\end{align}
where $ \mathbb{S}^i$ denotes the spin matrix of the atom at the position $\mathbf{R}^i$. 
One of the main properties of the spin operator is the commutation rule
\begin{align}
  [ \mathbb{S}_i,\mathbb{S}_j]=i \hbar \ze_{ijk}  \mathbb{S}_k  \; .
\end{align}
Here, $\ze$ denotes the antisymmetric Ricci tensor. The index in the spin operators refers to the component of the spin matrices along the cartesian axes. For a one-half spin particle, $s=1/2$ and $\mathbb{S}= \frac{\hbar}{2} \bs \zs $ where $\bs \zs $ are the vectors of the Pauli matrices. The value of the atomic spin  $\overline{\mathbf{S}^i} $ is obtained by taking the expectation value of the operator $ \mathbb{S}^i $ for the atomic spinorial wave function $\psi_i$.
\begin{align}
\overline{\mathbf{S}^i} \equiv  \langle\zy_i|  \mathbb{S}^i | \zy_i \rangle \;,  \label{expect S}
\end{align}
where we used the bracket Dirac notation. In particular, from Eq. \eqref{expect S} we have that the modulus of the atomic spin is  
\begin{align*}
  |\overline{\mathbf{S}^i}| = \hbar s_i
\end{align*}
as expected. The evolution of the spin expectation value is given by 
\begin{align}
   \dpp{\overline{\mathbf{S}^i}}{t} =- \frac{i}{\hbar}  \langle\zy_i|[ \mathbb{S}^i,\mathbb{H}_0]   | \zy_i \rangle = - \frac{g \zm_B}{\hbar} \overline{ \mathbf{S}^i} \times  \mathbf{B }_0\;,\label{Q evol S cons}
\end{align}
where we used the Schr\"odinger equation for the wave function  $\zy_i$
\begin{align}
 i\hbar \dpp{\zy_i}{t} &= \mathbb{H}_0 \zy_i\;. 
\end{align} 
It is useful to remark that the quantum Eq. \eqref{Q evol S cons} agrees with the classical evolution equation \eqref{evol S cons}. The evolution equation of a quantum spin in an external field is exactly the same of the classical case. The spin precesses around the field with a constant frequency that is independent from the value of the quantized spin $s$. In the following, we will see that this is also true for the binary exchange interaction. The only difference between the classical and the quantum spin evolution arises from the anisotropy field. %(?and the relaxation processes, other article?).

We describe now the evolution of the quantum system in a more general framework and we include the exchange and the anisotropy fields. The quantum evolution equation of a system can be obtained with two different strategies. The first one is the ``ab initio" approach: the analysis is focused on the elementary quantum mechanical  description of the particle dynamics. The relevant evolution equation is directly obtained by the Schr\"odinger or by some other equivalent formalism. In most cases,  the complexity of the mathematical formulation is reduced by taking some ad hoc approximations. The second possibility, that we will take in the following, is to start from the classical hamiltonian formulation of the dynamics. The quantum mechanical description is obtained by applying the so called quantization rules \cite{morandi_JPA_11,morandi_PRB_11b,morandi_PRA14}.
They provide a systematic procedure in which a function of some classical quantities (that in our case are the atomic spins $\mathbf{S}^i$) is associated to a suitable expression of the corresponding  quantum mechanical operators (in our case the atomic spin operators $ \mathbb{S}^i$).
The quantization procedure guarantees that the quantum mechanical dynamics is compatible with the classical dynamics in a suitable limit (the classical limit). For a spin system, the classical limit corresponds to the situation where the total spin $s$ of the atoms is very large 
\begin{align*}
\lim_{s_i\rightarrow \infty}   \langle \zy_i |  \mathbb{S}^i | \zy_i \rangle  & =   \mathbf{S}^i \; .
\end{align*} 
According to Eq. \eqref{Ham_cl} the classical Hamiltonian that describes the exchange and anisotropy fields is 
\begin{align}
 \mathcal{H}_{e-a} =   \sum_i      \mathcal{H}_{e-a}^i\;,
\end{align}
where
\begin{align}
 \mathcal{H}_{e-a}^i =     \sum_{  j\neq i, j\in \textrm{ NN}_i  } \frac{ J_{ij}}{  s_i s_j( g\mu_B)^2  } \left(   \mathbf{S}^i \cdot \mathbf{S}^j  \right)     -  \frac{  D}{ s_i^2 g\mu_B  }{S^i_z}^2\; . \label{Hea}
\end{align}
This expression shows that $\mathcal{H}_{e-a}^i$ is a quadratic form of the atomic spin vectors. In particular, the spin product in the first term refers to neighbor pairs. The quantification of the term $   \mathbf{S}^i \cdot \mathbf{S}^j   $ for $i\neq j$ is straightforward, it is sufficient to replace the classical vectors by the spin operators
\begin{align}
  \mathbf{S}^i \cdot  \mathbf{S}^j \rightarrow    \mathbb{S} ^i \cdot \mathbb{S} ^j \; .
\end{align}
The anisotropy term (the second term of Eq. \eqref{Hea}) is proportional to the squared value of the spin  at the position $ \mathbf{R}^i $. In this case the quantization rules are more complex (see for example Ref. \cite{Skomski_book}). %pag 31 panel 3
 We have
\begin{align}
 \mathbf{S}^i \cdot  \mathbf{S}^i \rightarrow 2 \; \mathbb{S}^i \cdot  \overline{\mathbf{S}^i}   -   \mathbb{S}^i \cdot \mathbb{S}^i  
\end{align}
With this identification, the find the quantum hamiltonian  
\begin{align}
 \mathbb{H}_{e-a}^i =       \sum_{  j\neq i, j\in \textrm{ NN}_i  } \frac{ J_{ij}}{  s_i s_j( g\mu_B)^2  } \left(   \mathbb{S} ^i \cdot \mathbb{S} ^j  \right)     -   \frac{  D}{ s_i^2 g\mu_B  } \left(   2 \; \mathbb{S}^i_z \cdot  \overline{ S^i_z}   -   {\mathbb{S}^i_z}^2   \right)\; .
\end{align}
The hamiltonian $\mathbb{H}_{e-a}^i$ acts on the many-body Hilbert space. In order to derive the Schr\"odinger equation for the $i-$th atoms, it is necessary to project $\mathbb{H}_{e-a}^i $ on the single particle Hilbert space. This can be done by taking the partial trace of $\mathbb{H}_{e-a}^i $ with respect all the atomic sites $\mathbf{R}^j$ with $j\neq i$. We obtain
\begin{align}
 \mathbb{H}_{e-a}^{ \textrm{sp},i} \equiv \textrm{tr}_{j\neq i} \left\{\mathbb{H}_{e-a}^i\right\} = &      \sum_{  j\neq i, j\in \textrm{ NN}_i  } \frac{ J_{ij}}{  s_i s_j( g\mu_B)^2  } \left(   \mathbb{S} ^i \cdot \overline{\mathbf{S}^j}  \right)     - \frac{  D}{ s_i^2 g\mu_B  } \left(   2 \; \mathbb{S}^i_z \cdot  \overline{ S^i_z}   -   {\mathbb{S}^i_z}^2   \right) \nn \\
 =  &  \frac{g \zm_B}{\hbar}   \mathbb{S} ^i \cdot  \mathbf{B}_{e-a}^i    -  \frac{  D}{ s_i^2 g\mu_B  }  {\mathbb{S}^i_z}^2\; ,
\end{align}
where we have defined 
\begin{align}
 \mathbf{B}_{e-a}^i=  &    \frac{1}{ g\mu_B\hbar} \sum_{  j\neq i, j\in \textrm{ NN}_i  } \frac{ J_{ij}}{  s_i s_j }     \overline{\mathbf{S}^j}   - \frac{  2 D}{ s_i^2 \hbar  }     \;    \overline{ S^i_z}  \widehat{\mathbf{e}}_z\; .
\end{align}
By adding the contribution of the external field we obtain the total  single particle Hamiltonian 
\begin{align}
 \mathbb{H}_{e-a}^{ \textrm{sp},i}+\mathbb{H}_0 &= \frac{g \zm_B}{\hbar} \mathbb{S} ^i \cdot \mathbf{B}^i     -  \frac{  D}{ s_i^2 g\mu_B  }  {\mathbb{S}^i_z}^2\;. 
\end{align}
It is easy to see that the Landau dumping process can be easily described with this formalism. Comparing Eq. \eqref{evol S cons} with Eq. \eqref{landau eq CL} we see that the inclusion of the Landau dumping mechanism to the precessional motion, is equivalent to substitute the external magnetic field with the following nonlinear effective field
\begin{align}
 \mathbf{B}_{L}^i &= \frac{1}{\za_i^2+1}\left(\mathbf{B}^i- \frac{\za_i}{s_i}\mathbf{S}^i \times  \mathbf{B}^i \right)\;. 
\end{align}
Proceeding in the same way, we obtain 
%\begin{align}
% \mathbb{H}(\mathbf{R}^i) &= \frac{g \zm_B}{\hbar(\za_i^2+1)}  \mathbb{S} ^i \cdot \left(\mathbf{B}+{\bs \xi} - \frac{\za_i}{s_i}\overline{\mathbf{S}^i} \times  \mathbf{B} \right)    -  \frac{  D}{ s_i^2 g\mu_B  }  {\mathbb{S}^i_z}^2
%\end{align}
\begin{align}
 \mathbb{H}(\mathbf{R}^i) &= \frac{g \zm_B}{\hbar(\za^2+1)}  \mathbb{S} \cdot \left(\mathbf{B}+{\bs \xi} - \frac{\za}{s}\overline{\mathbf{S}} \times  \mathbf{B} \right)    -  \frac{  D}{ s^2 g\mu_B  }  {\mathbb{S}_z}^2
\end{align}
where for simplicity we have dropped the index $i$ in the right side. The Hamiltonian $ \mathbb{H}^i$ describes the  non conservative LLG-Langevin spin dynamic in the quantum framework.  According to our previous discussion, it is easy to see that if we discard the last term of the equation the evolution of the spin expectation value $\overline{\mathbf{S}^i} \equiv  \langle\zy_i|  \mathbb{S}^i | \zy_i \rangle $ agrees with Eq. \eqref{landau eq CLnoise}.

\section{Conclusions}

In this paper we have discussed the possible extension of the Landau-Lifshitz-Gilbert theory to a quantum system. Our model consists of a array of quantum spin particles interacting via the exchange field. It is derived by applying the standard quantization rules to the LLG equation. The effects of an external thermal bath is reproduced by including a stochastic noise. As a final result, we show that by applying a direct quantization procedure the only measurable quantum effect arises from the anisotropy magnetic field.

\end{document}